\documentclass[article,accept,moreauthors,pdftex,10pt,a4paper,preprints]{mdpi} 

\firstpage{1} 

\history{}

\usepackage[utf8]{inputenc}
\usepackage[T1]{fontenc}
\usepackage{hyperref}
\usepackage{graphicx,bm,amsmath,color,scalerel}
\usepackage{bbold}
\usepackage{wasysym}
\usepackage{verbatim}

\newcommand{\be}{\begin{equation}}
\newcommand{\ee}{\end{equation}}
\newcommand{\beq}{\begin{eqnarray}}
\newcommand{\eeq}{\end{eqnarray}}
\newcommand{\ba}{\begin{align}}
\newcommand{\ea}{\end{align}}

\Title{Spacetime and deformations of special relativistic kinematics}

\Author{José Manuel Carmona$^{1}$*\orcidB{}, José Luis Cortés$^{1}$\orcidC{}, and José Javier Relancio$^{1}$\orcidD{}}

\AuthorNames{José Manuel Carmona, José Luis Cortés, and José Javier Relancio}

\address{%
$^{1}$ \quad Departamento de F\'{\i}sica Te\'orica and Centro de Astropartículas y Física de Altas Energías (CAPA),
Universidad de Zaragoza, Zaragoza 50009, Spain; jcarmona@unizar.es (J.M.C); cortes@unizar.es (J.L.C.); relancio@unizar.es (J.J.R.)}

\corres{Correspondence: jcarmona@unizar.es}

\abstract{A deformation of special relativistic kinematics (possible signal of a theory of quantum gravity at low energies) leads to a modification of the notion of spacetime. At the classical level, this modification is required when one considers a model including single- or multi-interaction processes, for which absolute locality in terms of canonical space-time coordinates is lost. We discuss the different alternatives for observable effects in the propagation of a particle over very large distances that emerge from the new notion of spacetime. A central ingredient in the discussion is the cluster decomposition principle, which can be used to favor some alternatives over the others.}


\begin{document}

\section{Introduction}
We start from the classical notion of spacetime based on the theory of Special Relativity (SR). The invariance of the equations of motion for different inertial reference frames related by Lorentz transformations (relativity principle) led Einstein to go from the notion of absolute time and space to the notion of spacetime in SR. Together with this notion of spacetime, one has the SR kinematics of particle processes: the energy of a particle is determined by its mass and the momentum of the particle (dispersion relation), and the invariance under spacetime translations implies energy-momentum conservation laws where the total energy-momentum of a system of free particles is the sum of the energy-momentum of each particle.   

At the quantum level, one finds that a relativistic quantum field theory is the quantum theory of relativistic particles with interactions compatible with SR kinematics and the property of cluster~\cite{Wichmann:1963aba}. For a pedagogical discussion of the derivation of relativistic quantum field theory as the only consistent framework for a relativistic quantum theory, see Ref.~\cite{Weinberg:1995mt}, where one can read:
\begin{itemize}
\item {\it ``... quantum field theory ... is the only way to reconcile the principles of quantum mechanics (including the cluster decomposition property) with those of special relativity.''} Steven Weinberg, The Quantum Theory of Fields, Preface to Volume I.
\item {\it ``It is one of the fundamental principles of physics (indeed, of all science) that experiments that are sufficiently separated in space have unrelated results... In S-matrix theory, the cluster decomposition principle states that if multiparticle processes .. are studied in ... very distant laboratories, then the S-matrix element for the overall process factorizes.''} Steven Weinberg, The Quantum Theory of Fields, Volume I, pg. 177.
\end{itemize}

Spacetime is introduced in the quantum theory through the combination of creation-annihilation operators required to implement the cluster decomposition principle (see, for example, Ref.~\cite{Maggiore:2005qv}). 
Einstein once more showed that if one wants at the classical level to include the gravitational interaction in a theory of relativistic particles, one has to go from special relativity to general relativity~\cite{Weinberg:1972kfs}. But at the quantum level we do not know how to include the gravitational interaction consistently in the framework of quantum field theory~\cite{Isham:1993ji}. We still do not have a well-defined quantum theory of gravity~\cite{Woodard:2009ns}.

The role of spacetime changes drastically when one goes from special relativity to general relativity. In special relativity  one has a fixed geometry but in general relativity the geometry of spacetime is a dynamical degree of freedom. This leads to consider the possibility that in the low-energy limit of a quantum theory of gravity the notion of spacetime and the related symmetries (Poincaré invariance) are deformed~\cite{Majid:1999tc}.

One can follow two different paths to explore this possibility. At the quantum level one can try to implement in a quantum field theory framework the  interaction of particles with a deformed relativistic kinematics and a possible deformed version of the cluster decomposition principle. A conjecture is that this will be associated to a noncommutative generalization of the notion of spacetime. This has led to explore the formulation of quantum field theory in a noncommutative spacetime. In the case of canonical noncommutativity (commutator of spacetime coordinates commuting with spacetime coordinates) one can follow the Weyl quantization method and give a systematic perturbative formulation of the noncommutative QFT~\cite{Szabo:2001kg,Minwalla:1999px,Seiberg:1999vs,Akofor:2008ae}. One finds a modification of the UV behaviour of the theory due to the noncommutativity of spacetime and a mixing IR/UV due to the nonlocality of interactions, while it is possible to find fully renormalizable models on Moyal space~\cite{Grosse:2004yu,Gurau:2005gd,Grosse:2009pa}.
In the case of a Lie algebra noncommutative spacetime ($\kappa$-Minkowski), there are new ambiguities when one tries to apply the Weyl quantization~\cite{Kosinski:1999ix,Grosse:2005iz,Meljanac:2010ps,Meljanac:2011cs,Juric:2015jxa,Juric:2015hda,Poulain:2018mcm,Juric:2018qdi,Poulain:2018two}, and the consistency of a formulation of a quantum field theory as well as the identification of the consequences of the noncommutativity of spacetime are open questions.  

The other path to explore the modification of the notion of spacetime is to implement at the classical level a deformation of the SR kinematics (modification of the energy-momentum relation and/or modification of the total energy-momentum of a system of free particles in terms of the energy-momentum of each particle) in a model for the interaction of particles and try to see which is the associated notion of spacetime in this framework. This is the path that we will follow in this work identifying the main obstacles and the different alternatives to overcome them.   

A particular objective of this study is to try to answer the question of whether there will be observable effects of deformation of the notion of spacetime when considering the propagation of particles over very large distances.

\section{Deformation of SR kinematics}

By SR kinematics we understand the restrictions on the momenta of the particles that participate in a process. We assume that these particles can be separated in two groups, in-state ($t\to -\infty$) and out-state ($t\to \infty$), and that, in each of these states, particles are separated by sufficiently large distances so that we can neglect any interaction between them and the total energy (momentum) of the in and out states is just the sum of the energies (momenta) of the particles in each state. A restriction on the momenta of the particles comes from the equality of the total momentum of the in and out state. The other restriction is obtained from the equality of the total energies of the in and out state when one uses the expression for the energy of a free particle ($E$) with a momentum $\vec{p}$ and a mass $m$, $E=\sqrt{ \vec{p}^2 + m^2}$. Those restrictions (conservation of momentum and energy) are a consequence of the space and time translational invariance.

We will consider a deformation of SR kinematics defined by a new scale ($\Lambda$) of energy or momentum (we work in units where $c= 1$ and then energy, momentum and mass have the same dimension). There are two ways to introduce the deformation of SR kinematics: 1) one modifies the expression of the energy of a particle with a given momentum to a $\Lambda$-dependent expression such that in the limit $(\vec{p}^2/\Lambda^2)\to 0$ one recovers the SR expression, 2) one modifies the expression of the total momentum (energy) of the in and out states in terms of the momenta (energies) of the particles in those states. We will refer to the two ways to introduce the deformation as deformed dispersion relation (DDR) and deformed composition law (DCL), respectively. In the general case one can consider simultaneously the two deformations. In fact, if one wants to preserve the invariance under (a deformed implementation of) Lorentz transformations, one has to combine the two deformations in an appropriate way. When the scale $\Lambda$ of the deformation is identified as a maximum energy or momentum for the possible states of a particle, one has what is known as double special relativity (DSR)~\cite{Amelino-Camelia2002,Kowalski-Glikman2005,AmelinoCamelia:2010pd,Bruno2001}.

When the deformation of the kinematics involves a DDR, but the standard undeformed additive composition law of momenta, a consequence of the introduction of a new scale $\Lambda$ in the expression of the energy of a particle in terms of its momentum is that the Lorentz invariance of SR is lost (Lorentz invariance violation, or LIV). One has a set of reference frames connected by rotations\footnote{We are assuming the energy is a function of the modulus of the momentum. A more general case would involve a dependence on the direction of the momentum. In this case one also looses the invariance under rotations.} (but not boosts) where the deformed ($\Lambda$-dependent) expression of the energy in terms of the momentum of the particle is valid. The conservation of momentum (energy) in SR kinematics is automatically incorporated in the quantum theory framework through locality (an action which is a space-time integral of a Lagrangian density at each space-time point depending on fields and derivatives of the fields at each point). When the deformation of SR kinematics does not involve a modification of the expression of the momentum (energy) of the in and out states in terms of the momenta (energies) of the particles, one can still have an implementation of the deformation in a local quantum theory framework~\cite{Colladay:1998fq}. In the LIV framework, the notion of spacetime in the quantum theory is not affected by the deformation of the kinematics.    

When the deformed kinematics involves a DCL, as in the case of DSR, the implementation of the deformation in a quantum framework will require to go beyond local quantum field theory, and the corresponding simple notion of spacetime is lost. Lacking a good candidate for the deformation of a local quantum field theory, we turn to the classical framework and try to extract a modified notion of spacetime from a formulation of a model for a process which implements a DCL.

In the derivation of relativistic quantum field theory, a crucial role was played by the cluster decomposition principle. At the classical level, this principle implies that processes that are sufficiently separated in space can be treated independently. In order to proceed with the identification of the notion of spacetime we have to be more specific about what we mean by a process.    

\section{Single-interaction process}

The simplest way, the one we advocate in this work, to define a process at the classical level is through one interaction involving all the particles in the in and out states.

We just consider interactions with no more than two particles in the initial state and, since at the classical level the particles in the final states are those of the initial state, then we have just one possible interaction with two particles in the initial and final states. We define the interaction through the conservation of the momentum and energy in the interaction. This leads to a model defined by the action
\begin{align}
S \,=& \int_{-\infty}^0 d\tau \sum_i \left[x_{-(i)}^\mu(\tau) \dot{p}_\mu^{-(i)}(\tau) + N_{-(i)}(\tau) \left[C(p_{-(i)}(\tau)) - m_{-(i)}^2\right]\right] \nonumber \\   
& + \int^{\infty}_0 d\tau \sum_j \left[x_{+(j)}^\mu(\tau) \dot{p}_\mu^{+(j)}(\tau) + N_{+(j)}(\tau) \left[C(p_{+(j)}(\tau)) - m_{+(j)}^2\right]\right] \nonumber \\
& + \xi^\mu \left[{\cal P}^+_\mu(0) - {\cal P}^-_\mu(0)\right]
\label{S1}
\end{align}
where $\dot{a}\doteq (da/d\tau)$ is a derivative of the variable $a$ with respect to the parameter $\tau$ along the trajectory of the particle, $x_{-(i)}$ ($x_{+(j)}$) are the spacetime coordinates of the in-state (out-state) particles, $p^{-(i)}$ ($p^{+(j)}$) their four-momenta, ${\cal P}^-$ (${\cal P}^+$) the total four-momentum of the in-state (out-state), $C(k)$ the function of a four-momentum $k$ defining the DDR, $\xi^\mu$ are Lagrange multipliers that implement the energy-momentum conservation in the interaction, and $N_{-(i)}$ ($N_{+(j)}$) Lagrange multipliers implementing the dispersion relation of in-state (out-state) particles.

The conditions for the action to be stationary are
\be
\dot{p}^{-(i)}_\mu \,=\, \dot{p}^{+(j)}_\mu \,=\, 0, \quad\quad
\frac{\dot{x}^\mu_{-(i)}}{N_{-(i)}} \,=\, \frac{\partial C(p^{-(i)})}{\partial p_\mu^{-(i)}}, \quad\quad \frac{\dot{x}^\mu_{+(j)}}{N_{+(j)}} \,=\, \frac{\partial C(p^{+(j)})}{\partial p_\mu^{+(j)}},
\ee
which is telling us that the momentum-energy of each particle is a constant along its trajectory and that there is a relation between the four-velocity and the four-momentum determined by the (deformed) dispersion relation. One also has
\be
{\cal P}^+_\mu \,=\, {\cal P}^-_\mu, \quad\quad\quad C(p^{-(i)}) \,=\, m_{-(i)}^2,
\quad\quad\quad C(p^{+(j)}) \,=\, m_{+(j)}^2,
\label{dk}
\ee   
which define the (deformed) kinematics, and    
\be
x_{-(i)}^\mu(0) \,=\, \xi^\nu \frac{\partial {\cal P}^-_\nu}{\partial p^{-(i)}_\mu}, \quad\quad\quad
x_{+(j)}^\mu(0) \,=\, \xi^\nu \frac{\partial {\cal P}^+_\nu}{\partial p^{+(j)}_\mu},
\ee
which fix the end (starting) spacetime coordinates of the trajectories of the in-state (out-state) particles. 

The solutions (trajectories of particles) are determined by a choice of constant four-momenta ($p^\mu_{-(i)}$, $p^\mu_{+(j)}$) compatible with the (deformed) kinematics equations (\ref{dk}) and the four constants $\xi^\mu$ (interaction vertex). The presence of the four arbitrary constants $\xi^\mu$ in the solutions is just a consequence of translational invariance.

When the deformation of the kinematics only affects to the dispersion relation one has
\be
{\cal P}^-_\mu \,=\, \sum_{i} p^{-(i)}_\mu, \quad\quad \quad
{\cal P}^+_\mu \,=\, \sum_{j} p^{+(j)}_\mu,     
\ee
and then one has
\be
x_{-(i)}^\mu(0) \,=\, x_{+(j)}^\mu(0) \,=\, \xi^\mu.
\ee
There is a crossing of the trajectories of all the particles at a point with spacetime coordinates $\xi^\mu$. The interaction defines a point in spacetime as in the case of SR.

If there is a DCL
\be
{\cal P}^-_\mu \,\neq\, \sum_{i} p^{-(i)}_\mu, \quad\quad \quad
{\cal P}^+_\mu \,\neq\, \sum_{j} p^{+(j)}_\mu,     
\ee
the trajectories of the particles do not cross and the interaction does not define a point in spacetime unless $\xi^\mu=0$; in this case, $x_{-(i)}^\mu(0)=x_{+(j)}^\mu(0)=0$ and all the trajectories cross at the origin. Different choices of $\xi^\mu$ just correspond to different observers related by translations. We have lost the locality of the interactions except for an observer whose origin coincides with interaction vertex~\cite{AmelinoCamelia:2010qv,AmelinoCamelia:2011bm}. This is in contrast to the case with no deformation of the composition law, where the interaction is local for all observers.   

One can ask what happend to the Lorentz invariance of SR kinematics after the introduction of the deformation. If the deformation does not introduce any direction in space (isotropic deformation) then in the general case we will just have a rotational invariance but the invariance under boosts is lost. But if the deformation in the dispersion relation (determined by $C(k)$) and the deformation in the composition law (expression of the total four-momentum in terms of the four-momenta of each particle) are appropriately chosen~\cite{Carmona2012a,Carmona2016} then one has relativistic deformed kinematics with observers related by Lorentz transformations having the same equations for the kinematics. The deformation requires that the boosts act nonlinearly on the four-momenta of the particles~\cite{Gubitosi:2013rna}. A simple example of a relativistic deformed kinematics is~\cite{Lukierski:2002df} 
\be
{\cal P}_0 \,=\, p^{(1)}_0 + p^{(2)}_0, \quad\quad {\cal P}_i \,=\, p_i^{(1)} + e^{-p_0^{(1)}/\Lambda} p_i^{(2)}, \quad\quad C(k) \,=\, \frac{\Lambda}{2} \left(e^{k_0/\Lambda} + e^{-k_0/\Lambda} - 2\right) - e^{k_0/\Lambda} \vec{k}^2.
\ee   

Let us go back to the issue of the loss of locality due to a DCL. This is telling us that the interaction is not defining a point in canonical spacetime (we are using phase space coordinates $x^\mu$, $p_\mu$ for each particle so that the term in the action corresponding to the propagation of a free particle is just $x^\mu\dot{p}_\mu$). In Ref.~\cite{Carmona2018}, the question of whether it is possible to find new spacetime coordinates such that the interaction is local (all particles have the same new spacetime coordinates at the interaction $\tau=0$) is addressed with an afirmative answer. The new spacetime coordinates are
\be
\tilde{x}^\alpha_{(1)} \,=\, x^\mu_{(1)} \varphi^\alpha_\mu(p^{(1)}) + x^\mu_{(2)} \varphi^{(2)\alpha}_{(1)\mu}(p^{(2)}), \quad\quad
\tilde{x}^\alpha_{(2)} \,=\, x^\mu_{(2)} \varphi^\alpha_\mu(p^{(2)}) + x^\mu_{(1)} \varphi^{(1)\alpha}_{(2)\mu}(p^{(1)}),
\ee
with functions $\varphi^\alpha_\mu(k)$, $\varphi^{(2)\alpha}_{(1)\mu}(k)$ and $\varphi^{(1)\alpha}_{(2)\mu}(k)$ determined by the DCL. From this result one concludes that the interaction defines new (correlated) spacetime coordinates ($\tilde{x}_{(1)}$, $\tilde{x}_{(2)}$) for a two-particle system which differ from the canonical spacetime coordinates ($x_{(1)}$, $x_{(2)}$) due to the deformation of the composition of momenta. When the (modulus of the) components of one of the four-momenta (for example $p^{(2)}$) are much smaller than the scale of deformation $\Lambda$, then the new spacetime coordinates of the other particle reduce to $\tilde{x}^\alpha_{(1)} = x^\mu_{(1)} \varphi^\alpha_\mu(p^{(1)})$. The correlation is lost but the new spacetime coordinates still differ from the canonical spacetime coordinates ($x_{(1)}$).      

We can now consider the consistency of the classical model with a deformed kinematics with the cluster decomposition principle. We have implicitly used this principle when we considered the process with just one interaction, independently of the interactions which would produce the particles in the in-state and the interactions used to detect the particles in the out-state. This is justified by the large distances of these interactions with respect to the single-interaction of the process. But the identification of the coordinates $\tilde{x}$ as the physical spacetime coordinates implies a correlation of the particles even when they are separated by large distances. Then the propagation of one particle depends on the four-momentum of a particle which is separated by a very large distance. This is not consistent with the assumption that processes separated by large distances can be treated independently. In fact, if we consider the propagation of a free particle as a process then we should be able to treat the propagation of the particles in the in-state (or those in the out-state) independently when they become separated by very large distances. Then the expressions for the correlated two-particle space-time coordinates ($\tilde{x}_{(1)}$, $\tilde{x}_{(2)}$) should apply at short distances from the interaction vertex but there should be a transition from this correlated two-particle space-time coordinates to a system of uncorrelated space-time coordinates. This transition where the correlation is lost is still compatible with a modification of the notion of spacetime due to the deformation of the kinematics if we assume that the physical space-time coordinates for a particle separated by large distances from any other particle are
\be
\tilde{x}^\alpha \,=\, x^\mu \varphi^\alpha_\mu(p).
\label{1p-phys-st}
\ee
This will be the assumption we will use to study the possible observable effects of a deformation of SR kinematics in the propagation of a particle over very large distances.

\section{Multi-interaction process}

In Ref.~\cite{AmelinoCamelia:2011nt}, a multi-interaction process was considered, defined as a transition between an in-going and an out-going state involving several interactions. Each interaction combines some of the particles in the in-going and out-going states together with particles propagating between different interactions.

To avoid irrelevant notation complications we just consider a process with two two-particle interactions which transform a three-particle in-going state with phase coordinates $(x_{-(i)}, p^{-(i)})$  onto a three-particle out-going state with phase space coordinates $(x_{+(j)}, p^{+(j)})$. Only two of the three in-state particles ($i=1,2$) participate in the first interaction, and only two of the out-state particles in the second interaction ($j=2,3$). There is one additional particle, with phase space coordinates $(y, q)$, which is produced in the first interaction and participates in the second interaction. The action which generalizes the action considered in the single-interaction process (\ref{S1}) is
\begin{align}
  S \,=& \int_{-\infty}^{\tau_1} d\tau \sum_{i=1,2} \left[x_{-(i)}^\mu(\tau) \dot{p}_\mu^{-(i)}(\tau) + N_{-(i)}(\tau) \left[C(p_{-(i)}(\tau)) - m_{-(i)}^2\right]\right] \nonumber \\
  & + \int_{-\infty}^{\tau_2} \left[x_{-(3)}^\mu(\tau) \dot{p}_\mu^{-(3)}(\tau) + N_{-(3)}(\tau) \left[C(p_{-(3)}(\tau)) - m_{-(3)}^2\right]\right] \nonumber \\   & + \int_{\tau_1}^{\tau_2} d\tau \left[y^\mu(\tau) \dot{q}_\mu(\tau) + N(\tau) \left[C(q(\tau)) - m^2\right]\right] \nonumber \\ 
& + \int_{\tau_1}^{\infty} \left[x_{+(1)}^\mu(\tau) \dot{p}_\mu^{+(1)}(\tau) + N_{+(1)}(\tau) \left[C(p_{+(1)}(\tau)) - m_{+(1)}^2\right]\right] \nonumber \\ & +   \int_{\tau_2}^{\infty} d\tau \sum_{j=2,3} \left[x_{+(j)}^\mu(\tau) \dot{p}_\mu^{+(j)}(\tau) + N_{+(j)}(\tau) \left[C(p_{+(j)}(\tau)) - m_{+(j)}^2\right]\right] \nonumber \\
& + \xi^\mu \left[\left(p^{+(1)}\oplus q\oplus p^{-(3)}\right)_\mu - \left(p^{-(1)}\oplus p^{-(2)}\oplus p^{-(3)}\right)_\mu\right](\tau_1) \nonumber \\ & + \chi^\mu \left[\left(p^{+(1)}\oplus p^{+(2)}\oplus p^{+(3)}\right)_\mu - \left(p^{+(1)}\oplus q\oplus p^{-(3)}\right)_\mu\right](\tau_2),
\end{align}
where we have introduced the notation $(k\oplus p\oplus q)$ for the total four-momentum of a system of three particles with four-momenta $(k, p, q)$.

The condition for the action to be stationary under changes $\delta x^\mu(\tau)$ and $\delta p_\mu(\tau)$ leads to
\be
\dot{p}^{-(i)} \,=\, \dot{p}^{+(j)} \,=\, \dot{q} \,=\, 0, \quad\quad
\frac{\dot{x}_{-(i)}^\mu}{N_{-(i)}} \,=\, \frac{\partial C(p^{-(i)})}{\partial p^{-(i)}_\mu}, \quad\quad \frac{\dot{x}^\mu_{+(j)}}{N_{+(j)}} \,=\, \frac{\partial C(p^{+(j)})}{\partial p_\mu^{+(j)}}, \quad\quad \frac{\dot{y}^\mu}{N} \,=\, \frac{\partial C(q)}{\partial q_\mu}.
\label{eq:velocity}
\ee
The deformed kinematics is defined by
\begin{align}
  & C(p^{-(i)}) \,=\, m_{-(i)}^2, \quad\quad C(p^{+(j)}) \,=\, m_{+(j)}^2, \quad\quad C(q^2) \,=\, m^2, \nonumber \\
  & p^{-(1)}\oplus p^{-(2)}\oplus p^{-(3)} \,=\, p^{+(1)}\oplus q\oplus p^{-(3)} \,=\, p^{+(1)}\oplus p^{+(2)} \oplus p^{+(3)},
\label{dk2}
\end{align}
which are derived when one considers the variations $\delta N_-$, 
$\delta N_+$,  $\delta N$, $\delta \xi^\mu$, $\delta\chi^\mu$,
and, from the variations $\delta p_\mu(\tau_1)$, $\delta q_\mu(\tau_1)$,
$\delta p_\mu(\tau_2)$, $\delta q_\mu(\tau_2)$, one also has
\begin{align}
  & x^\mu_{-(i)}(\tau_1) \,=\, \xi^\nu \frac{\partial(p^{-(1)}\oplus p^{-(2)}\oplus p^{-(3)})_\nu}{\partial p^{-(i)}_\mu}, \:\: (i=1,2) \quad\quad x^\mu_{-(3)}(\tau_2) \,=\, \chi^\nu \frac{\partial(p^{+(1)}\oplus q\oplus p^{-(3)})_\nu}{\partial p^{-(3)}_\mu}, \nonumber \\
  & x^\mu_{+(1)}(\tau_1) \,=\, \xi^\nu \frac{\partial(p^{+(1)}\oplus q\oplus p^{-(3)})_\nu}{\partial p^{+(1)}_\mu}, \quad\quad x_{+(j)}^\mu(\tau_2) \,=\, \chi^\nu \frac{\partial(p^{+(1)}\oplus p^{+(2)} \oplus p^{+(3)})_\nu}{\partial p^{+(j)}_\mu}, \:\: (j=2,3), \nonumber \\
  & y^\mu(\tau_1) \,=\, \xi^\nu \frac{\partial(p^{+(1)}\oplus q\oplus p^{-(3)})_\nu}{\partial q_\mu}, \quad\quad y^\mu(\tau_2) \,=\, \chi^\nu \frac{\partial(p^{+(1)}\oplus q\oplus p^{-(3)})_\nu}{\partial q_\mu}.
\label{eq:endpoints}
\end{align}
The equations of the kinematics allow to determine the four-momentum $q$ and give restrictions between the momenta of the in-state and out-state particles as in the case of the single-interaction process. One also has relations between the four-velocities of the in-state and out-state particles and their momenta. 

The new ingredient due to the presence of two interactions is that, on the one hand, from equation~\eqref{eq:velocity} for the four-velocity of the particle propagating between the two interactions, and using $\dot{q}=0$, one has
\be
y^\mu(\tau_2) - y^\mu(\tau_1) \,=\, \int_{\tau_1}^{\tau_2} d\tau\, \dot{y}^\mu(\tau) \,=\, \frac{\partial C(q)}{\partial q_\mu} \,\int_{\tau_1}^{\tau_2} d\tau\, N(\tau),
\ee
and from equations~\eqref{eq:endpoints} for the end points of the trajectory of the particle propagating between the two interactions,
\be
 y^\mu(\tau_2) - y^\mu(\tau_1) \,=\, \left(\chi^\nu - \xi^\nu\right) \,\frac{\partial(p^{+(1)}\oplus q\oplus p^{-(3)})_\nu}{\partial q_\mu}.
\ee
Combining both equations, we get a relation for the coordinates of the two vertices,
\be
\left(\chi^\nu - \xi^\nu\right) \,\frac{\partial(p^{+(1)}\oplus q\oplus p^{-(3)})_\nu}{\partial q_\mu} \,=\, \frac{\partial C(q)}{\partial q_\mu} \,\int_{\tau_1}^{\tau_2} d\tau N(\tau).
\ee
Then, although there are two vertices, the difference of coordinates of the two vertices is fixed and one has a set of solutions depending on four arbitrary constants ($\xi^\mu$) as in the case of the single-interaction process, which reflects the invariance under translations. 

There is one observer (for which $\xi^\mu=0$) that sees the first interaction as local but not the second one, and another observer related by a translation (for which $\chi^\mu=0$) seeing the second interaction as local but not the first one. For any other observer both interactions are not local. This is what is known as relative locality. 

In a recent work~\cite{Gubitosi:2019ymi}, the possibility to have a non-linear implementation of Lorentz transformations connecting the different solutions of the two-interaction process has been worked out in detail. 

\section{Observable effects in the propagation of a particle over very large distances}
\label{sec:observable}

We do not have any signal of a deformation of SR kinematics in laboratory experiments. This can be due to the small values of the ratios $\vec{p}^2/\Lambda^2$ due to our limitations to have very high energy particles and leads to the necessity to look for amplification mechanisms of the effects of the deformation. Besides the use of the extreme sensitivity of reaction thresholds to the SR kinematics, which has been used to put bounds on the scale of deformation when one considers a modification of the dispersion relation~\cite{Mattingly:2005re}, another possibility is to consider the propagation of particles over very large distances assuming that one can have corrections proportional to the distance leading to observable effects. In order to explore this possibility, a model for the effects of a deformation of the kinematics in the propagation of a particle is needed. We consider three alternatives:
\begin{enumerate}  

\item \emph{Model for the propagation based on a DDR.} The simplest model for the propagation of a particle is based on neglecting the interaction term in the action (\ref{S1}): the extrema of the action correspond to the solutions of   
\be
\dot{p}_\mu \,=\, 0, \quad\quad\quad \frac{\dot{x}^\mu}{N} \,=\, \frac{\partial C(p)}{\partial p_\mu}.
\ee
The velocity ($\vec{v}$) of propagation of a particle with a given momentum $\vec{p}$ is 
\be
v^i \,=\, \frac{\partial C(p)/\partial p_i}{\partial C(p)/\partial p_0},
\ee
where $p_0$ is the positive solution of the equation $C(p)=m^2$. The dependence of the velocity on the momentum $\vec{p}$ will be modified with respect to the standard SR relation due to the DDR (deformed dispersion relations appear naturally in the context of $\kappa$-Minkowski spacetime, see e.g. Ref.~\cite{Juric:2015jxa}). In this model for the propagation of the particle, there is no place to effects due to the DCL. The model will lead to a dependence of the velocity of propagation of photons ($m=0$) on the energy, which will have consequences on the spectral time distribution. Spectral and timing observations from active galactic nuclei (AGN) and short gamma-ray bursts (GRB) have been used to get bounds~\cite{Albert:2007qk,Martinez:2008ki,Ackermann:2009aa,HESS:2011aa,Nemiroff:2011fk,Vasileiou:2013vra,Vasileiou:2015wja} on the scale of deformation $\Lambda$ of the order of the Planck scale, although some analyses cast doubts on such stringent bounds~\cite{Amelino-Camelia:2017zva,Xu:2018ien}.

\bigskip
\bigskip

\item \emph{Model for the propagation based on a DDR and a physical spacetime defined by the DCL.} A less trivial model for the propagation of a particle is based on the observation that spacetime coordinates should be defined by the crossing of worldlines in the interaction of particles. This, together with the decoupling of particles at very large distances due to the cluster decomposition principle, leads to identify the linear combinations of the canonical spacetime coordinates with coefficients depending on momentum variables 
\be
\tilde{x}^\alpha \,=\, x^\mu \varphi^\alpha_\mu(p)
\ee
as the physical spacetime coordinates. In this case one has a velocity of propagation for a particle with a given momentum
\be
\tilde{v}^i \,=\, \frac{\dot{\tilde{x}}^i}{\dot{\tilde{x}}^0} \,=\, \frac{\dot{x}^\mu \varphi^i_\mu(p)}{\dot{x}^\nu \varphi^0_\nu(p)} \,=\, \frac{(\partial C(p)/\partial p_\mu) \varphi^i_\mu(p)}{(\partial C(p)/\partial p_\nu) \varphi^0_\nu(p)}.
\ee
The dependence of the velocity on the momentum is determined by the DDR (through $C(p)$) and also by the DCL (through $\varphi^\alpha_\mu(p)$, which is determined by the DCL). Both ingredients of the deformation of SR kinematics have to be considered in the propagation of a particle.     
 
The relation between the DCL and $\varphi^\alpha_\mu(p)$ has been determined to be~\cite{Carmona2018,Carmona:2019fwf}
\be
\varphi^\alpha_\mu(p) \,=\, \lim_{q\to 0} \frac{\partial(q\oplus p)_\mu}{\partial q_\alpha}.
\ee 
An example of a DDR and DCL leading to a relativistic deformed kinematics is~\cite{Lukierski:2002df} 
\be
C(p) \,=\, \Lambda^2 \left(e^{p_0/\Lambda} + e^{-p_0/\Lambda} - 2\right) - e^{p_0/\Lambda} \vec{p}^2, \quad\quad (q\oplus p)_0 \,=\, q_0 + p_0, \quad\quad (q\oplus q)_i \,=\, q_i + e^{-q_0/\Lambda} p_i,
\label{bcb}
\ee
 so that 
\begin{align}
& \frac{\partial C(p)}{\partial p_0} \,=\, \Lambda \left(e^{p_0/\Lambda} - e^{-p_0/\Lambda}\right) - e^{p_0/\Lambda} \vec{p}^2/\Lambda, \quad\quad \frac{\partial C(p)}{\partial p_i} \,=\, - 2 e^{p_0/\Lambda} p_i, \nonumber \\
& \varphi^0_0(p) \,=\, 1, \quad\quad \varphi^i_0(p) \,=\, 0, \quad\quad \varphi^0_i(p) \,=\, - p_i/\Lambda, \quad\quad \varphi^i_j(p) \,=\, \delta^i_j,
\end{align}
and then 
\be
\tilde{v}^i \,=\, \frac{- 2 e^{p_0/\Lambda} p_i}{\Lambda \left(e^{p_0/\Lambda} - e^{-p_0/\Lambda}\right) + e^{-p_0/\Lambda} \vec{p}^2/\Lambda}.
\ee 
The dispersion relation $C(p)=m^2$ gives
\be
e^{p_0/\Lambda} \vec{p}^2/\Lambda^2 \,=\, e^{p_0/\Lambda} + e^{-p_0/\Lambda} - 2 - m^2/\Lambda^2,
\ee
and the result for the energy dependence of the velocity is
\be
1 - \vec{\tilde{v}}^2 \,=\, \frac{m^2 \left(1 + m^2/4\Lambda^2\right)}{\left[\Lambda\left(e^{p_0/\Lambda} -1\right) - m^2/2\Lambda\right]^2},
\label{v(E)}
\ee
generalizing the SR result $1-v^2 = m^2/p_0^2$. There is no effect of the deformation in the  velocity of propagation of photons ($m=0$) and the first order correction in an expansion in powers of the inverse of the deformation scale for a massive particle is
\be
1 - \vec{\tilde{v}}^2 \approx \frac{m^2}{p_0^2} \left(1 - \frac{p_0}{\Lambda}\right),
\ee
which will have consequences on the timing of different signals from an astrophysical transient. 

We have not considered, in all the discussion of the model for the effects of a deformation of the kinematics in the propagation of a particle, the arbitrariness in the starting point corresponding to the choice of canonical coordinates in phase space. In fact if one considers new momentum coordinates $p'_\mu$ related nonlinearly to $p_\nu$ then one will have a new dispersion relation defined by a function $C'$ and a new modified composition law $\oplus'$ which are related to the function $C$ and $\oplus$ by
\be
C(p) \,=\, C'(p'), \quad\quad\quad (q'\oplus' p')_\mu \,=\, (q\oplus p)'_\mu \,.
\ee  
Then we have 
\be
\begin{split}
\varphi'^\alpha_\mu(p') &\,=\, \lim_{q'\to 0} \frac{\partial(q'\oplus' p')_\mu}{\partial q'_\alpha} \,=\, \lim_{q'\to 0} \frac{\partial(q\oplus p)'_\mu}{\partial q'_\alpha} \,=\, \lim_{q\to 0} \frac{\partial q_\beta}{\partial q'_\alpha} \frac{\partial(q\oplus p)'_\mu}{\partial q_\beta}\\ & \,=\, \lim_{q\to 0} \frac{\partial(q\oplus p)'_\mu}{\partial(q\oplus p)_\nu} \frac{\partial(q\oplus p)_\nu}{\partial q_\alpha} \,=\, \frac{\partial p'_\mu}{\partial p_\nu} \varphi^\alpha_\nu(p).
\end{split}
\ee
On the other hand, the nonlinear change of momentum variables $p\to p'$ defines a canonical change of coordinates in phase space with 
\be
x'^\mu \,=\, x^\rho \frac{\partial p_\rho}{\partial p'_\mu}\,,
\ee
and then 
\be
x'^\mu \varphi'^\alpha_\mu(p') \,=\, x^\rho \frac{\partial p_\rho}{\partial p'_\mu} \varphi'^\alpha_\mu(p') \,=\, x^\rho \frac{\partial p_\rho}{\partial p'_\mu} \frac{\partial p'_\mu}{\partial p_\nu} \varphi^\alpha_\nu(p) \,=\, x^\nu \varphi^\alpha_\nu(p).
\ee
This means that $\tilde{x}'^\alpha=\tilde{x}^\alpha$ and then one has $\tilde{v}'^i(p'_0)=\tilde{v}^i(p_0)$. Every relativistic deformed kinematics obtained from (\ref{bcb}) by a canonical transformation of coordinates of the form $p\to p'=f(p)$ (what is usually called a change of `basis' in momentum space) will not show any effect of the deformation in the propagation of photons, and the modified energy dependence of the velocity for a massive particle can be obtained directly from (\ref{v(E)}) by using the nonlinear change of momentum variables together with the dispersion relation to express $p_0$ in terms of $p'_0$.  

The absence of an energy dependence for the velocity of propagation of photons in the model for the propagation of a particle with a relativistic deformed kinematics has not taken into account the expansion of the universe. But this can not be neglected in the propagation over astrophysical distances required to have a possible observable effect. It is an open question how to combine consistenly a deformation of SR kinematics with the cosmological model based on a curved spacetime. Although some effects have been considered in the propagation of particles in a curved $\kappa$-Minkowski spacetime~\cite{Harikumar:2012zi}, there are different approaches to include a curvature in spacetime in a deformed relativistic kinematics. This has been done, in the DSR framework, using Finsler spacetimes for curved spacetimes~\cite{Letizia:2016lew}. A different approach was used in Ref.~\cite{Barcaroli:2015xda}, where the modification is carried out by Hamiltonian geometry (see also Ref.~\cite{2012arXiv1203.4101M}). In this case, the metric is momentum dependent (the Hamiltonian version of a Lagrange space). The starting point in all of them is a modified dispersion relation, but a key ingredient of a modified relativistic kinematics is a deformed composition law, which does not appear in the previous works. Another approach is considered in Ref.~\cite{Cianfrani:2014fia}, where the authors try to combine a curvature in momentum space and in spacetime including a modified composition law, generalizing the original relative locality action introducing what they called non-local variables. In any case, it remains to be seen whether the absence of signals of the deformed kinematics in timing measurements applies after the proper inclusion of the expansion of the universe.

\bigskip
\bigskip

\item \emph{Model for the propagation based on a DDR, including the interactions at production and detection of the particle that determine the initial and final points of the trajectory through the DCL.} The study of the two-interaction process in the previous section suggests to use the results for the trajectory ($y^\mu(\tau)$) of the particle between the two interactions as a third alternative model for the effects of a deformation of relativistic kinematics on the propagation of a particle identifying the two interactions in the process with the production and detection of the particle. This model has in common with the first simplest model based on neglecting the interaction term in the action (\ref{S1}) that the velocity of propagation of the particle $v^i \,=\, \dot{y}^i/\dot{y}^0 \,=\, (\partial C(q)/\partial q_i)/(\partial C(q)/\partial q_0)$ will have a momentum dependence determined by the DDR. However, in order to determine the time of flight, one has to consider that the end points of the trajectory of the particles ($y^\mu(\tau_1)$, $y^\mu(\tau_2)$), and then the length of the path of the particle, depend also on the momentum of the particle and the momenta of other particles participating in the production and detection of the particles. One will have an spectral and timing distribution of gamma-rays from a short GRB which differs from the expectation in SR but one does not have a definite prediction on the effect of the deformation of SR kinematics due to its dependence on details of the production and detection interactions to which we do not have access. On top of these problems, this third model for the effects of a deformed kinematics on the propagation of a particle based on a process with two interactions presents doubts on its consistency. One has to justify why one can consider the process with just two interactions. One can ask about the interactions  in the production of each of the three particles in the in-state and the detection of each of the three particles in the out-state. The argument to neglect these interactions is that they are separated by very large distances from the two interactions in the process but if we want to use the model to study the effect of a deformation of SR kinematics on the propagation of a particle over large distances then the two interactions are already separated by a very large distance and then, according with the implementation of the cluster decomposition principle at the classical level, we should treat the process with two interactions as two independent single-interaction processes. The conclusion of this discussion is that the third model for the propagation of a particle is disfavoured with respect to the other two models.      
\end{enumerate}

Having discussed the three alternatives for a model for the propagation of particles, we can try to answer the question whether there will be observable effects of the deformation of the kinematics through the distorsion of the temporal-spectral distribution of emision and detection of gamma rays from a given source. Concentrating on the second alternative, the naive answer would be that there is no distortion since photons of different energies propagate with the same velocity, but the problem is not so simple. One has to consider two observers, one whose spatial origin is within the source (whose size is negligible compared with the distance between the source and the detector) and another one within the detector. In order to determine the path of each photon from the emission to the detection it is necessary to identify the transformation (translation) between the two observers. 

From the action for the propagation of a photon
\be
S \,=\, \int d\tau \left[\dot{x}^\mu p_\mu - N(\tau) C(p)\right] \,=\, \int d\tau \left[-\tilde{x}^\alpha \varphi^\mu_\alpha(p) \dot{p}_\mu - N(\tau) C(p)\right],
\ee
one would identify translations with the transformations $x^\mu \to x^\mu + a^\mu$ which leave the action invariant. But one can see that the transformation $\tilde{x}^\mu \to \tilde{x}^\mu + a^\mu$, which is not a symmetry of the action, is a transformation that leaves invariant the set of equations \eqref{eq:velocity}-\eqref{eq:endpoints} (and then the set of their solutions). There are differences in some equations (as it should be since the action is not invariant) but the differences cancel when the rest of equations are taken into account. 

There are then two alternatives for the translation between observers. If one takes the first option (the symmetry of the action) then one has a momentum dependent transformation of the spacetime coordinates $\tilde{x}$ and one concludes that even if the velocity of propagation of photons of different energies is the same in this spacetime, the translation between observers leads generically to a time delay (there are bases in which there is no time delay). This is the perspective which is adopted for the spacetime in the relative locality proposal~\cite{AmelinoCamelia:2011cv,Loret:2014uia}. On the other hand, if one takes the second option for the translation between observers, the independence on the energy of the velocity of propagation of photons leads to the absence of time delays. This is the perspective of spacetime that we advocate in this work. In a future work we will show how the second alternative for translations emerge from a derivation of the effective model for the propagation of a photon from a model which includes the interactions at the source which produce the photons and the interactions in the detector.

\section{Conclusions and outlook}

In this work we have taken as starting point the assumption that a low energy signal of a quantum theory of gravity will be a deformation of SR kinematics based on a classical spacetime. We have discussed from different perspectives how to introduce a modified notion of spacetime associated to a deformed kinematics. 

A crucial ingredient in the discussion is the cluster decomposition principle which is at the basis of making possible to do physics with an isolated system. We have presented different classical models corresponding to different identification of spacetime coordinates and to different implementations of the cluster decomposition principle. We have given arguments in favor of one of the options which leads to a model for the propagation of particles over very large distances such that when the deformation of SR kinematics is compatible with Lorentz invariance one finds that the velocity of propagation of photons turns out to be independent of the energy and there are no time delays. While, as indicated in Sec.~\ref{sec:observable}, the experimental situation is not yet clear, we have left open the question of whether this result will still be valid after incorporating the expansion of the Universe. 


\authorcontributions{All authors contributed equally to the present work.}

\acknowledgments{This work is supported by Spanish grants PGC2018-095328-B-I00 (FEDER/Agencia estatal de investigación), and DGIID-DGA No. 2015-E24/2.
The authors would also like to thank support from the COST Action CA18108, and acknowledge Giulia Gubitosi, Flavio Mercati and specially Giacomo Rosati for useful conversations.}

\conflictsofinterest{The authors declare no conflict of interest.} 

\externalbibliography{yes}
\bibliography{QuGraPhenoBib}
\end{document}